\documentclass[aps,prl,twocolumn,superscriptaddress,nofootinbib]{revtex4-1}
\usepackage{amsmath}
\usepackage{amssymb}
\usepackage{graphicx}
\usepackage{mathrsfs}
\usepackage{ntheorem}
\usepackage{mathtools}
\usepackage{enumerate}
\usepackage{enumitem}
\usepackage{times,txfonts}
\usepackage{subfigure}
\usepackage{upgreek}
\usepackage{tikz}
\newcommand{\ket}[1]{|#1\rangle}
\newcommand{\bra}[1]{\langle #1|}
\newcommand{\Tr}{\mathrm{Tr}}

\newcommand{\abs}[1]{\lvert #1\rvert}

\def\CC{{\rm\kern.24em \vrule width.04em height1.46ex depth-.07ex \kern-.30em C}}
\def\RR{{\rm\kern.24em \vrule width.04em height1.46ex depth-.07ex
\kern-.30em R}}
\def\P{{\rm I\kern-.25em P}}

\begin{document}

\title{Deterministic Coherence Distillation}

\author{C. L. Liu}

\affiliation{Institute of Physics, Beijing National Laboratory for
  Condensed Matter Physics, Chinese Academy of Sciences, Beijing
  100190, China}

\author{D. L. Zhou}

\email{zhoudl72@iphy.ac.cn}

\affiliation{Institute of Physics, Beijing National Laboratory for
  Condensed Matter Physics, Chinese Academy of Sciences, Beijing
  100190, China}

\affiliation{School of Physical Sciences, University of Chinese
  Academy of Sciences, Beijing 100049, China}

\affiliation{CAS Central of Excellence in Topological Quantum
  Computation, Beijing 100190, China}

\affiliation{Songshan Lake Materials Laboratory, Dongguan, Guangdong
  523808, China}

\date{\today}

\begin{abstract}
  Coherence distillation is one of the central problems in the
  resource theory of coherence. In this Letter, we complete the
  deterministic distillation of quantum coherence for a finite number
  of coherent states under strictly incoherent operations.
  Specifically, we find the necessary and sufficient condition for the
  transformation from a mixed coherent state into a pure state via
  strictly incoherent operations, which recovers a connection between
  the resource theory of coherence and the algebraic theory of
  majorization lattice. With the help of this condition, we present
  the deterministic coherence distillation scheme and derive the
  maximum number of maximally coherent states obtained via this
  scheme.
\end{abstract}

\maketitle

\emph{Introduction.}-- Quantum coherence is a valuable resource in
performing quantum information processing tasks \cite{Nielsen}. It can
implement various information processing tasks that cannot be
accomplished classically, such as quantum computing
\cite{Shor,Grover}, quantum cryptography \cite{Bennett}, quantum
metrology \cite{Giovannetti,Giovannetti1}, and quantum biology
\cite{Lambert}. Recently, the resource theory of coherence has
attracted a growing interest due to the development of quantum
information science
\cite{Aberg,Baumgratz,Chitambar,Chitambar1,Fan,Levi,Winter,
Yadin,Yu,Vicente,Streltsov}.

All quantum resource theories have two fundamental ingredients: free
states and free operations \cite{Brandao,Liu1}. For the resource theory of
coherence, the free states are the quantum states that are diagonal in a
prefixed reference basis. However, there is no general consensus on
the set of free operations. Based on different physical and
mathematical considerations, a number of free operations were proposed
\cite{Aberg,Baumgratz,Chitambar,Chitambar1,Winter,Yadin}. Here, we
focus our discussion on the strictly incoherent operations. This type
of free operation was first given in Ref. \cite{Winter} and was shown
that it can neither create nor use coherence and has a physical
interpretation in terms of interferometry in Ref. \cite{Yadin}. Thus,
the strictly incoherent operations are a physically well-motivated set
of free operations for coherence and a strong candidate for free operations.

One of the central problems in the resource theory of coherence is the
coherence distillation
\cite{Baumgratz,Lami,Lami1,Yuan,Winter,Liu,Fang,Zhao,Brandao,
  Regula,Regula1,Bu,Chitambar2}, which is the process that extracts
pure coherent states from general states via free operations. This
problem was approached in two different settings: the asymptotic
regime \cite{Winter,Brandao,Yuan,Lami,Lami1,Chitambar2,Bu} and
the one-shot regime \cite{Liu,Zhao,Regula,Regula1}. Although many
interesting results have been obtained, however, there are still some
open fundamental questions remaining to be solved. One of which is the
deterministic coherence distillation, whose aim is to find the
condition of conversion from a general mixed state to the maximally
coherent state with certainty \cite{Zhao1,Streltsov,Regula1}.
Investigations on this topic have been started in Ref. \cite{Regula1},
where the deterministic coherence distillation of pure coherent states
under several classes of incoherent operations was introduced.
However,  the deterministic coherence
distillation of general mixed states has been left as an open question.

In this Letter, we address the above question by completing the
framework for deterministic coherence distillation under strictly
incoherent operations. We first recall some notions of the resource
theory of coherence and the notions of majorization
lattice which are related to our topic. Then, we present the necessary
and sufficient condition for the transformation from a general state
into a pure state via strictly incoherent operations, which recovers a
connection between the resource theory of coherence and the algebraic
theory of majorization lattice. With the help of this condition, we
present the deterministic coherence distillation scheme. Then, we derive
the maximum number of maximally coherent states that can be obtained
in this deterministic coherence distillation scheme.

\emph{Resource theory of coherence.}--Let $\mathcal{H}$ represent the
Hilbert space of a $d$-dimensional quantum system. A particular basis
of $\mathcal{H}$ is denoted as $\{\ket{i}, ~i=0,1,\cdots,d-1\}$, which is
chosen according to the physical problem under discussion.
Specifically, a state is said to be incoherent if it is diagonal in
the basis. We represent the set of incoherent states as $\mathcal{I}$.
Any state that cannot be written as a diagonal matrix is defined as a
coherent state. Note that the term coherent state here is different from
 the canonical coherent state or the spin coherent state \cite{Nielsen}. For a pure state 
$\ket{\varphi}$, we will denote $\ket{\varphi}\bra{\varphi}$ as $\varphi$, i.e.,
$\varphi:=\ket{\varphi}\bra{\varphi}$ and we will denote
$\ket{\varphi_m^d}=\frac1{\sqrt{d}}\sum_{i=0}^{d-1}\ket{i}$ as a
$d$-dimensional maximally coherent state.

A strictly incoherent operation is a completely positive
trace-preserving map, expressed as
$\Lambda(\rho)=\sum_n K_n\rho K_n^\dagger$, where the Kraus operators
$K_n$ satisfy not only $\sum_n K_n^\dagger K_n= I$ but also
$K_n\mathcal{I}K_n^\dagger\subset \mathcal{I}$ and
$K_n^\dagger\mathcal{I}K_n\subset\mathcal{I}$ for $K_n$, i.e., each
$K_n$ as well as $K_n^\dagger$ maps an incoherent state to an incoherent state.
With this definition, it is elementary to show that a projector is an
incoherent operator if and only if it has the form
$\mathbb{P}_{\text{I}}=\sum_{i\in{\text{I}}}\ket{i}\bra{i}$ with
$\text{I}\subset\{0,1,...,d-1\}$. In what follows, we will denote
$\mathbb{P}_{\text{I}}$ as strictly incoherent projective operators. The
the dephasing map, which we will denote as $\Delta(\cdot)$, is defined as
$\Delta\rho=\sum_{i=0}^{d-1}\ket{i}\bra{i}\rho\ket{i}\bra{i}$.

\emph{Majorization and majorization lattice.}-- Majorization
\cite{Bhatia} is a mathematical tool widely used in quantum
information theory \cite{Nielsen1,Du,Zhu}. For the $n$-dimensional probability
distributions $\mathcal{P}^n$, we say that a probability distribution
$\textbf{p}=(p_1,p_2,...,p_n)$ is majorized by
$\textbf{q}=(q_1,q_2,...,q_n)$, in symbols
$\textbf{p}\prec\textbf{q}$, if there are
$\sum_{i=1}^lp_i^\downarrow\leq\sum_{i=1}^lq_i^\downarrow$, for all
$1\leq l\leq n$, where $\downarrow$ indicates that the elements are to be taken in
descending order. The majorization lattice \cite{Cicalese,Davey,Yu1} is a
quadruple $(\mathcal{P}^n,\prec,\vee,\wedge)$. Here $\prec$ is the relation introduced
above. For every pair of $\textbf{p},\textbf{q}\in\mathcal{P}^n$,
$\textbf{p}\wedge\textbf{q}$ is the unique greatest lower bound of
$\textbf{p},\textbf{q}$ up to a permutation transformation which is defined as a probability
distribution, for every $\textbf{s}\in\mathcal{P}^n$ with
$\textbf{s}\prec\textbf{p}$, $\textbf{s}\prec\textbf{q}$, then there is
$\textbf{s}\prec\textbf{p}\wedge\textbf{q}$; and
$\textbf{p}\vee\textbf{q}$ is the unique least upper bound of
$\textbf{p},\textbf{q}$ which is defined as a probability distribution
for every $\textbf{t}\in\mathcal{P}^n$ with
$\textbf{p}\prec\textbf{t}$ and $\textbf{q}\prec\textbf{t}$, then there is
$\textbf{p}\vee\textbf{q}\prec\textbf{t}$. Similarly, we write
$\bigwedge \mathcal{S}$ as the unique greatest lower bound of
$\mathcal{S}$ and $\bigvee\mathcal{S}$ as the unique least upper bound of
$\mathcal{S}$, where $\mathcal{S}$ is a subset of $\mathcal{P}^n$.
 Hereafter, we will apply majorization to density operators
and write $\rho_1\prec\rho_2$ if and only if the corresponding majorization relation holds
for the eigenvalues of $\rho_1$ and $\rho_2$. And $\bigvee\mathcal{S}\prec\rho$
means that the least upper bound (up to a unitary transformation) of
$\mathcal{S}$ is majorized by $\rho$.


\emph{Determined state transformation.}--In the following, we will give
the necessary and sufficient condition for a state $\rho$ can be
transformed into a pure coherent state $\ket{\varphi}$ via strictly incoherent
operations.

\emph{Theorem 1.} We can transform a mixed state $\rho$ into a pure coherent state
$\varphi$ via strictly incoherent operations if and only if there exists an
orthogonal and complete set of incoherent projectors
$\{\mathbb{P}_\alpha\}$ such that, for all $\alpha$, there are
\begin{eqnarray}
  \frac{\mathbb{P}_{\alpha}\rho\mathbb{P}_{\alpha}}{\Tr(\mathbb{P}_{\alpha}\rho\mathbb{P}_{\alpha})}
  =\psi_{\alpha}~~\text{and}~~\Delta\psi_{\alpha}\prec\Delta\varphi, \label{Theorem_1}
\end{eqnarray}
where $\psi_\alpha$ are all pure coherent states. In other words, there exists
$\{\mathbb{P}_\alpha\}$ such that
\begin{eqnarray}
  \bigvee\mathcal{S}\prec\Delta\varphi,
\end{eqnarray}
where $\mathcal{S}$ is the set of $\{\Delta\psi_\alpha\}$.

\emph{Proof.} First, we show that $\rho$ can be transformed into
$\varphi$ via a strictly incoherent operation if and only if
$P\rho P^t$ (superscript $t$ means transpose) can be transformed 
into $\varphi$ via a strictly incoherent operation with $P$ being a 
permutation matrix.

For any two strictly incoherent operations $\Lambda_1$ with Kraus operators
$\{K_n^1\}$ and $\Lambda_2$ with Kraus operators $\{K_m^2\}$, the operation
 $\Lambda=\Lambda_1\circ\Lambda_2$ is also a strictly incoherent
operation with Kraus operators $\{K_l=K_n^1K_m^2\}$,
since we can easily verify it by examining
$K_l\mathcal{I}K_l^\dag\subseteq\mathcal {I}$ and
$K_l^\dag\mathcal{I}K_l\subseteq\mathcal{I}$.
It is straightforward to verify that, for any permutation matrix, both
$P$  and its inverse are strictly incoherent operations. With these knowledge, it is easy to show that $\rho$ can be transformed into
$\varphi$ via a strictly incoherent operation if and only if
$P\rho P^t$ can be transformed into $\varphi$ via a strictly incoherent
operation. Hence, without loss of generality, we let
\begin{eqnarray}
  \rho=\bigoplus_\mu p_\mu\rho_\mu, \label{direct_sum}
\end{eqnarray}
corresponding to the Hilbert space
$\mathcal{H}=\bigoplus_\mu\mathcal{H}_\mu$ with each $\rho_\mu$ being irreducible. Here,
an irreducible matrix $\rho_\mu$ means that it cannot be transformed into a block diagonal matrix by using a permutation matrix.

Second, we show the \emph{if} part of the theorem, i.e., if the state
$\rho$ satisfies the condition in the theorem above, then we can
transform a mixed state $\rho$ into a pure state $\varphi$ via a strictly
incoherent operation.

Let $\rho$ be a state satisfying the condition in the theorem above.
Then, according to the result in Ref. \cite{Chitambar,Zhu,Du} which says that
a pure coherent state $\ket{\psi}$ can be transformed into another pure coherent state
$\ket{\varphi}$ via strictly incoherent operations if and only if there is
$\Delta\psi\prec\Delta\varphi$, we can always find strictly incoherent operations
$\Lambda_{\alpha}(\cdot)$, which act on the support of
$\mathbb{P}_{\alpha}$, with $\Lambda_{\alpha}(\cdot)=\sum_n K_{\alpha}^n(\cdot){K_{\alpha}^n}^\dag$, such that
\begin{eqnarray}
  \Lambda_\alpha(\psi_{\alpha})=\varphi, \nonumber
\end{eqnarray}
for all $\alpha$. With this result, we  transform $\rho$ into
$\ket{\varphi}$ by using the operation
\begin{eqnarray}
  \Lambda(\cdot)=\bigoplus_{\alpha}\Lambda_{\alpha}(\cdot), \nonumber
\end{eqnarray}
where the corresponding Kraus operators are
\begin{eqnarray}
  K_{\alpha,n}=K_{\alpha}^n\oplus \textbf{0}. \nonumber
\end{eqnarray}
Here, $\textbf{0}$ represents a square matrix with all its elements
being zero. It is straightforward to show that $\Lambda(\cdot)$ is a
strictly incoherent operation.

Third, we show the \emph{only if} part of the theorem, i.e., if $\varphi$ can
be obtained from a state $\rho$ via a strictly incoherent operation, then
the state $\rho$ should satisfy the condition in the theorem above.

Let us assume that we can obtain a pure coherent state $\varphi$ from a mixed state
$\rho$ by using a strictly incoherent operation $\Lambda(\cdot)$. Then, there is
\begin{eqnarray}
  \Lambda(\rho)=\sum_nK_n\rho K_n^\dag=\varphi. \label{kraus_k}
\end{eqnarray}
Substituting Eq. (\ref{direct_sum}) into (\ref{kraus_k}), we can
obtain that
\begin{eqnarray}
  \Lambda(\rho)=\sum_{n,\mu}p_\mu K_n\rho_\mu K_n^\dag=\varphi. \label{channel_redirect}
\end{eqnarray}
Since pure states are extreme points of the set of states, there must
be
\begin{eqnarray}
  K_n\rho_\mu K_n^\dag=q_{n,\mu}\varphi, \label{rho_alpha} \nonumber
\end{eqnarray}
for all $n$ and $\mu$, where $q_{n,\mu}=\Tr(K_n\rho_\mu K_n^\dag)$.

According to the definition of the strictly incoherent operations,  there 
is at most one nonzero element in each column (row) of a
strictly incoherent Kraus operator. Thus, any $K_n$ can always be
decomposed into
\begin{eqnarray}
  K_n=P_\pi K_n^D\mathbb{P}_n, \label{Kraus_decomposion}
\end{eqnarray}
where the operator $P_\pi$ is a permutation matrix,
$K_n^D=\text{diag}(a_1,...,a_n,0,0,...)$ is a diagonal matrix with
$a_i$ being nonzero complex numbers, and $\mathbb{P}_n$ is a
projective operator corresponding to $K_n^D$, i.e.,
$\mathbb{P}_n=\text{diag}(1,...,1,0,0,...)$. Let
$\{p_{\mu,i},\ket{\psi_{\mu,i}}\}$ be an arbitrary ensemble decomposition of
$\rho_\mu$. Then, there is
\begin{eqnarray}
  K_n\rho_\mu K_n^\dag=\sum_{\mu,i}p_{\mu,i}P_\pi K_n^D\mathbb{P}_n\psi_{\mu,i} \mathbb{P}_n{K_n^D}^\dag P_\pi^\dag.\label{channel_afdirect}
\end{eqnarray}
From Eqs. (\ref{channel_redirect}) and (\ref{channel_afdirect}), we
 obtain that
\begin{eqnarray}
  \Lambda(\rho)=\sum_{n,\mu,i}p_\mu p_{\mu,i}P_\pi K_n^D\mathbb{P}_n \psi_{\mu,i} \mathbb{P}_n{K_n^D}^\dag P_\pi^\dag=\varphi. \nonumber
\end{eqnarray}
Again, by using the fact that pure states are extreme points of the
set of states, we immediately obtain that
\begin{eqnarray}
  \frac{P_\pi K_n^D\mathbb{P}_n \psi_{\mu,i} \mathbb{P}_n{K_n^D}^\dag P_\pi^\dag}{\Tr(P_\pi K_n^D\mathbb{P}_n \psi_{\mu,i} \mathbb{P}_n{K_n^D}^\dag P_\pi^\dag)}=\varphi~~ \text{or}~~\textbf{0}, \label{Kraus_identity}
\end{eqnarray}
for all $\mu$, $i$, and $n$. Clearly, $\ket{\psi_{\mu,i}}$ are states of the subspace
$\mathcal{H}_\mu$. Thus, we only need to consider the projective
operator $\mathbb{P}_n$ in Eq. (\ref{Kraus_decomposion}) corresponding
to the subspace $\mathcal{H}_\mu$ and we denoted it as
$\mathbb{P}_{n,\mu}$. Since $\Lambda$ is a trace preserving map, we can get
that $\sum_nK_n^\dag K_n=I$ and, furthermore,
$\sum_n\mathbb{P}_{n,\mu}K_n^\dag K_n\mathbb{P}_{n,\mu}=I_\mu$ with
$I_\mu$ being the identity matrix of the subspace $\mathcal{H}_\mu$. Here,
since every $\rho_\mu$ is irreducible,
$P_\pi K_n^D\mathbb{P}_n \ket{\psi_{\mu,i}}$ cannot be
a zero vector at the same time.

From Eq. (\ref{Kraus_identity}) and
$\sum_n\mathbb{P}_{n,\mu}K_n^\dag K_n\mathbb{P}_{n,\mu}=I_\mu$, we  get that
\begin{eqnarray}
  \mathbb{P}_{n,\mu}\psi_{\mu,i}\mathbb{P}_{n,\mu}=
  \mathbb{P}_{n,\mu}\psi_{\mu,j}\mathbb{P}_{n,\mu}~~\text{or}~~\textbf{0}, \label{projective_identity}
\end{eqnarray}
for all $i$ and $j$. Both these two cases mean that
$\frac{\mathbb{P}_{n,\mu}\rho_\mu\mathbb{P}_{n,\mu}}{\Tr(\mathbb{P}_{n,\mu}\rho_\mu\mathbb{P}_{n,\mu})}$
is a pure coherent state and we denoted it as $\psi_{n,\mu}$ for the sake of
simplicity. By using the condition that $\Lambda(\rho)=\varphi$ and the condition in
Eq. (\ref{projective_identity}), we immediately derive that
\begin{eqnarray}
  \Lambda(\psi_{n,\mu})=\varphi, \nonumber
\end{eqnarray}
for every $n$ and $\mu$. Since the state $\psi_{n,\mu}$ can be transformed
into $\varphi$ via a strictly incoherent operation if and only if
$\Delta\psi_{n,\mu}\prec\Delta\varphi$, we immediately obtain the conclusion in our theorem.
This completes the proof of the \emph{only if} part. ~~~~~~~~~~~~~~~~~~~~~~~~~~~~~~~~~~$\square$

From Theorem 1, we  infer the following corollary:

\emph{Corollary}. We can transform $\rho$ into a pure coherent state
$\psi$ via strictly incoherent operations if and only if $\psi_\alpha$ are all
coherent states for some $\{\mathbb{P}_\alpha\}$. 

\emph{Proof.} The \emph{only if} part follows directly from Theorem 1. To prove the \emph{if} part, 
without loss of generality, let us assume that
$\ket{\psi_\alpha}=\sum_{i=1}^{d_\alpha}c_i^\alpha\ket{i}$ with the number of
$c_i^\alpha>0$ being $d_\alpha\geq2$, and
$c_1^\alpha\geq\cdots\geq c_{d_\alpha}^\alpha$. From the definition of the majorization lattice,
we can immediately obtain that
$\bigvee\mathcal{S}\prec\bigvee\mathcal{S}^\prime$, where
$\mathcal{S}^\prime=\{\Delta\psi_\alpha^\prime\}$ with
$\ket{\psi_\alpha^\prime}=c_1^\alpha\ket{1}+\sum_{i=2}^{d_\alpha}c_i^\alpha\ket{i}$. Noting that the
set $\mathcal{S}^\prime$ is an ordered set \cite{Bhatia} and
$c_1^\alpha<1$, we then obtain that $\bigvee\mathcal{S}^\prime$ equals to one of
$\Delta\psi_\alpha^\prime$ and this corresponds to a coherent state
$\ket{\psi}$ where $\ket{\psi}=c_1\ket{1}+c_2\ket{2}$ with $0<c_1<1$.  ~~~~~~~~~~~~~~~~~~~~~~~~$\square$

\emph{Deterministic coherence distillation.}--Next, let us move to the
deterministic coherence distillation of a finite number of coherent
states.

Suppose that we have $n$ coherent states
\begin{eqnarray}
  \rho_1,\rho_2,...,\rho_n, \nonumber
\end{eqnarray}
where $\rho_1,\rho_2,...,\rho_n$ are not necessarily identical and $n$ is a
finite number. The deterministic coherence distillation process is the
process that extracts pure coherent states from them with certainty.
Here, we concentrate our discussion on the task that extracts as more
$2$-dimensional maximally coherent state $\ket{\varphi_m^2}$ as possible
from $\rho_1\otimes\rho_2\otimes\cdots\otimes\rho_n$ via strictly incoherent operations.

Based on the result above, we  take the distillation
procedure as the following three steps (See Fig.1).
\\
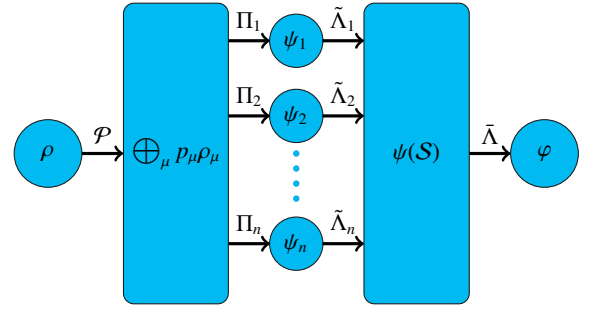
\begin{figure}[ht]
\begin{tikzpicture}
	[
	L1Node/.style={circle,draw=black!90,fill=cyan!80, thin, minimum size=9mm},
	L2Node/.style={rectangle,rounded corners,draw=black!90,fill=cyan!80, thin, minimum width = 0.8cm, minimum height = 4.0cm, text centered},
	L3Node/.style={rectangle,rounded corners,draw=black!90,fill=cyan!80,thin, minimum width = 1.4cm, minimum height = 4.0cm, text centered},
	L4Node/.style={circle,draw=black!90,fill=cyan!80, thin, minimum size=9mm},
	L5Node/.style={circle,draw=black!90,fill=cyan!80,thin, minimum size=4mm},	
      L6Node/.style={circle,draw=black!90,fill=cyan!80,thin, minimum size=4mm},	
      L7Node/.style={circle,draw=black!90,fill=cyan!80,thin, minimum size=4mm}]
	\node[L1Node] (n1) at (0, 0){$\rho$};
      \node[L2Node] (n2) at (1.7, 0){$\bigoplus_{\mu}p_\mu\rho_\mu$};
	\node[L5Node] (n5) at (3.3, 1.5){$\psi_1$};
	\node[L6Node] (n6) at (3.3, 0.5){$\psi_2$};
	\node[L7Node] (n7) at (3.3, -1.2){$\psi_n$};
	\node[L3Node] (n3) at (4.9, 0){$\psi(\mathcal{S})$};
	\node[L4Node] (n4) at (6.6, 0){$\varphi$};
	\draw [color=black!100,very thick,->] (0.45, 0)--node[anchor=south] {$\mathcal{P}$} (1, 0);
	\draw [color=black!100,very thick,->] (2.4,1.5)--node[anchor=south] {$\Pi_1$} (2.95, 1.5);
	\draw [color=black!100,very thick,->] (2.4, 0.5)--node[anchor=south] {$\Pi_2$} (2.95, 0.5);
	\draw [color=black!100,very thick,->] (2.4, -1.2)--node[anchor=south] {$\Pi_n$} (2.95, -1.2);
	\draw [color=black!100,very thick,->] (3.65,1.5)--node[anchor=south] {$\tilde{\Lambda}_1$} (4.2,1.5);
	\draw [color=black!100,very thick,->] (3.65,0.5)--node[anchor=south] {$\tilde{\Lambda}_2$} (4.2,0.5);
	\draw [color=black!100,very thick,->] (3.65,-1.2)--node[anchor=south] {$\tilde{\Lambda}_n$} (4.2,-1.2);
	\draw [color=black!100,very thick,->] (5.6,0)--node[anchor=south] {$\bar{\Lambda}$} (6.15,0);
	\filldraw [fill=cyan!90,draw=cyan!90] (3.3,0) circle [radius=1pt];
	\filldraw [fill=cyan!90,draw=cyan!90] (3.3,-0.2) circle [radius=1pt];
	\filldraw [fill=cyan!90,draw=cyan!90] (3.3,-0.4) circle [radius=1pt];
	\filldraw [fill=cyan!90,draw=cyan!90] (3.3,-0.6) circle [radius=1pt];
\end{tikzpicture}
 \caption{Schematic picture of the deterministic coherence transformation via strictly incoherent
 operations. Here, $\Pi_\alpha=\mathbb{P}_\alpha\cdot\mathbb{P}_\alpha$ for incoherent projective operator
 $\mathbb{P}_\alpha$, $\psi(\mathcal{S})$ is the pure coherent state determined by $\bigvee\mathcal{S}$,
 $\tilde{\Lambda}_\alpha$ are the strictly incoherent operations such that
  $\tilde{\Lambda}_\alpha(\psi_\alpha)=\psi$, $\bar{\Lambda}$ is the strictly incoherent operation
   such that $\bar{\Lambda}(\psi)=\varphi$, and all the others are the same as in the main text.}
\end{figure}
\\
First, for the given
$\rho=\rho_1\otimes\rho_2\otimes\cdots\otimes\rho_n$, we should transform
$\rho$ into a block diagonal matrix.

To this end, one should calculate out the permutation matrix $P$ that
can transform $\rho$ into a block diagonal matrix, i.e., the permutation
matrix $P$ such that
\begin{eqnarray}
  \mathcal{P}(\rho)=P\rho P^t=\bigoplus_{\mu=1}^Lp_\mu\rho_\mu\bigoplus\textbf{0}, \label{rho_permutation}
\end{eqnarray}
where each $\rho_\mu=\sum_{i,j}\rho_{ij}^\mu\ket{i}\bra{j} $
($\mu=1,2,\cdots,n$) is an irreducible density operator defined on the
$d_\mu$-dimensional subspace $\mathcal {H}_\mu$, $p_\mu>0$ satisfies
$\sum_{\mu=1}^Lp_\mu=1$, and $\textbf{0}$ represents a square matrix of
dimension $d_0=d-\sum_{\mu=1}^Ld_\mu$ with all its elements being zero.

Second, we should calculate out an incoherent projective operators set
$\{\mathbb{P}_\alpha\}$ in Theorem 1.

To this end, let us first introduce the following three matrices,
which are useful to obtain the corresponding $\{\mathbb{P}_\alpha\}$. For
$\rho=\sum_{ij}\rho_{ij}\ket{i}\bra{j}$, we can define two matrices
$|\rho|$ and $(\Delta\rho)^{-\frac12}$, where $|\rho|$ reads
$|\rho|=\sum_{ij}|\rho_{ij}|\ket{i}\bra{j}$ and
$(\Delta\rho)^{-\frac12}$ is a diagonal matrix with elements
$$
(\Delta\rho)^{-\frac12}_{ii}= \left\{
  \begin{array}{ll}
    \rho_{ii}^{-\frac12}, &\text{if} ~ \rho_{ii}\neq0;\\
    0,&\text{if}~ \rho_{ii}= 0.
  \end{array}\right.
$$
Next, we recall the following matrix with the help of $|\rho|$ and
$(\Delta\rho)^{-\frac12}$ 
\begin{eqnarray}
  \mathcal{A}=(\Delta\rho)^{-\frac12}\abs{\rho}(\Delta\rho)^{-\frac12}. \label{A}
\end{eqnarray}
A useful property of $\mathcal{A}$ is that all the elements of
$\mathcal{A}$ are $1$ if and only if $\rho$ is a pure coherent state \cite{Liu}.
By substituting the expression in Eq. (\ref{rho_permutation}) into Eq.
(\ref{A}), we  obtain that
\begin{eqnarray}
  \mathcal{A}=(\Delta\rho)^{-\frac12}\abs{\rho}(\Delta\rho)^{-\frac12}=
  \bigoplus_{\mu=1}^L\mathcal{A}_\mu\bigoplus\textbf{0 },\nonumber
\end{eqnarray}
where
$\mathcal{A}_\mu=(\Delta\rho_\mu)^{-\frac12}\abs{\rho_\mu}(\Delta\rho_\mu)^{-\frac12}$ are also
irreducible nonnegative matrices. Next, we should find out all the maximally
dimensional principal submatrices $\mathcal{A}_\mu^n$ of
$\mathcal{A}_\mu$ with all its elements being 1, where the maximal
dimension means that the dimension of $\mathcal{A}_\mu^n$ cannot be
enlarged. Let the corresponding Hilbert subspaces of principal
submatrices $\mathcal{A}_\mu^n$ be $\mathcal{H}_\mu^n$ spanned by
$\{\ket{i_\mu^1},\ket{i_\mu^2},\cdots,\ket{i_\mu^{d_n}}\}\subset\{\ket{0},
\ket{1},\cdots,\ket{d-1}\}$. Then, the corresponding incoherent projective
operators are
\begin{eqnarray}
  \mathbb{P}_\alpha=\ket{i_\mu^1}\bra{i_\mu^1}+\ket{i_\mu^2}\bra{i_\mu^2}+
\cdots+\ket{i_\mu^{d_n}}\bra{i_\mu^{d_n}}.\nonumber
\end{eqnarray}
Performing $\{\mathbb{P}_\alpha\}$ on the state $\rho$, we  obtain
$\{\psi_\alpha\}$, i.e.,
\begin{eqnarray}
  \frac{\mathbb{P}_{\alpha}\rho\mathbb{P}_{\alpha}}{\Tr(\mathbb{P}_{\alpha}\rho\mathbb{P}_{\alpha})}
  =\psi_\alpha.\nonumber
\end{eqnarray}
By the way, we  note that the set of $\{\mathbb{P}_\alpha\}$ in
Theorem 1 is not necessarily unique, and we denote the set of
$\{\Delta\psi_\alpha\}$ corresponding to the maximally dimensional principal
submatrices $\mathcal{A}_\mu^n$ as $\mathcal{S}_m$.

Third, we should calculate out the least upper bound of the set
$\mathcal{S}_m=\{\Delta\psi_{\alpha}\}$, i.e., $\bigvee\mathcal{S}_m$.

Without loss of generality, suppose that
$\ket{\psi_\alpha}=\sum_{i=1}^{d_n}c_\alpha^i\ket{i}$ and the corresponding
probability distributions of $\ket{\psi_\alpha}$ are
$\textbf{p}_{\alpha}^\downarrow=(\abs{c_\alpha^1}^2,\abs{c_\alpha^2}^2,...,\abs{c_\alpha^{d_n}}^2,0,0,...0)$.
Let us show how to calculate out the least upper bound of
$\mathcal{S}_m$, i.e., $\bigvee\mathcal{S}_m$. To this end, we first define
a probability distribution $\textbf{a}=(a_1,a_2,...,a_d)$, where
\begin{eqnarray}
  a_i=\max\{\sum_{j=1}^i\abs{c_1^j}^2,\sum_{j=1}^i\abs{c_2^j}^2,...,
\sum_{j=1}^i\abs{c_L^j}^2\}-\sum_{j=1}^{i-1}a_j.\nonumber
\end{eqnarray}
We  note that the elements of $\textbf{a}=(a_1,a_2,...,a_d)$
might not be in nonincreasing order, i.e., it is not true in general
that $a_j\geq a_{j+1}$. Apart from $\textbf{a}$, we also need the
following lemma, which was proved in Ref. \cite{Cicalese}.

\emph{Lemma}. Let $\textbf{a}=(a_1,a_2,...a_d)$ be a given probability
distribution, and let $j$ be the smallest integer in $\{2,...,n\}$
such that $a_j>a_{j-1}$. Moreover, let $i$ be the greatest integer in
$\{1,2,...,j-1\}$ such that
$a_{i-1}\geq\frac{\sum_{r=i}^ja_r}{j-i+1}=a$. Let the probability
distribution $\textbf{q}=(q_1, q_2,...,q_d)$ be defined as
\begin{equation}
  q_r=\left\{
    \begin{array}{ll} a, &\text{for} ~ r=i,i+1,...,j;\\
      a_r,&\text{otherwise}.
    \end{array}
  \right.\nonumber
\end{equation}
Then for the probability distribution $\textbf{q}$, we have that
$ q_{r-1}\geq q_r,~~\text{for~all}~r=2,...,j$, and
$\sum_{s=1}^kq_s\geq\sum_{s=1}^ka_s, ~~k=1,...,d$. Moreover, for all
$\textbf{t}=(t_1,t_2,...,t_d)$ such that
$\sum_{s=1}^kt_s\geq\sum_{s=1}^ka_s,~~k=1,...,n$, we also have
$\sum_{s=1}^kt_s\geq\sum_{s=1}^kq_s,~~k=1,...,n$.

By using the definition of $\textbf{a}$ and the iterate application of
the above Lemma, we can obtain the least upper bound of
$\mathcal{S}_m=\{\Delta\psi_\alpha\}$, i.e., $\bigvee\mathcal{S}_m$ and we denoted it as
$\Delta\psi$.

Without loss of generality, let the maximum number of $\varphi_m^2$ we can
distill from $\rho_1\otimes\rho_2\otimes\cdots\otimes\rho_n$ be $N$. The generalization to
$d>2$ is straightforward. From Theorem 1, this distillation can be
accomplished if the following majorization relation holds:
\begin{eqnarray}
  \Delta\psi\prec\text{diag}(2^{-N},...,2^{-N},0...,0). \label{majorization_maximal}
\end{eqnarray}
The above relation can be fulfilled if and only if
\begin{eqnarray}
  \|\psi\|_\infty\leq 2^{-N}, \label{majorization_first}
\end{eqnarray}
where $\|\cdot\|_\infty$ is the max norm on the matrix space. This can be
examined directly since if the first inequality of majorization
relation in Eq. (\ref{majorization_maximal}) holds, then the other
inequalities for Eq. (\ref{majorization_maximal}) are automatically
satisfied.

Thus, the inequality in Eq. (\ref{majorization_first}) gives the
maximum number of $2$-dimensional maximally coherent state that can be
distilled from $\rho_1\otimes\rho_2\otimes\cdots\otimes\rho_n$ and the maximum number is
\begin{eqnarray}
  N_{\max}=\lfloor\log_2\|\psi\|_\infty^{-1}\rfloor,\nonumber
\end{eqnarray}
where $\lfloor x\rfloor$ represents the largest integer equal to or less than $x$.

We can then summarize the above results as Theorem 2.

\emph{Theorem 2.} The maximum number of $2$-dimensional maximally
coherent state that can  distill from a set of states, such as
$\rho_1,\rho_2,...,\rho_n$, is
\begin{eqnarray}
  N_{\max}=\lfloor\log_2\|\psi\|_\infty^{-1}\rfloor.
\end{eqnarray}

In particular, if the states we chose are all pure coherent states
$\{\ket{\varphi_\gamma}\}$ with $\gamma=1,...,n$, then the maximum number of
$2$-dimensional maximally coherent state that we can be distilled is
$N_{\max}=\lfloor\log_2\otimes_{\gamma=1}^n\|\varphi_\gamma\|_\infty^{-1}\rfloor$, which corresponds to the
result in \cite{Regula1}. This is reminiscent of the case of
entanglement \cite{Nielsen1,Morikoshi,Hayashi}, where the
deterministic entanglement distillation of pure entangled states was
studied.

We should note that there is a class of
states that cannot be distilled into any pure coherent state via strictly
incoherent operations. If we can transform
$\rho=\sum_{ij}\rho_{ij}\ket{i}\bra{j}$ with the number of
$\rho_{ii}\neq0$ being $m$ into a pure coherent state
$\ket{\varphi}=\sum_ic_i\ket{i}$ with the number of $c_{i}\neq0$ being
$n$ via a strictly incoherent operation, then the rank of $\rho$ is at
most $\frac mn$. To see this, suppose that we can distill a pure
coherent state $\varphi$ from $\rho$, according to Theorem 1, there must be an
orthogonal and complete set of incoherent projectors
$\{\mathbb{P}_\alpha\}$ fulfilling the condition in Eq. (\ref{Theorem_1}).
Let the corresponding decomposition of the Hilbert space of
$\{\mathbb{P}_\alpha\}$ be
$\mathcal{H}=\bigoplus_\alpha\mathcal{H}_\alpha$, where the dimension of
$\mathcal{H}_\alpha$ is $d_\alpha$, the projections
$\{\mathbb{P}_\alpha\}$ of $\rho$ onto each $\mathcal{H}_\alpha$ are
$\{\psi_\alpha\}$, respectively, and
$\rho=\sum_{i=1}^l\uplambda_i\ket{\uplambda_i}\bra{ \uplambda_i}$ is a
spectral decomposition for $\rho$. Then, there are
\begin{eqnarray}
  \frac{\mathbb{P}_{\alpha}\ket{\uplambda_i}\bra{ \uplambda_i}\mathbb{P}
_{\alpha}}{\Tr(\mathbb{P}_{\alpha}\ket{\uplambda_i}\bra{ \uplambda_i}\mathbb{P}_{\alpha})}
  =\psi_{\alpha},\nonumber
\end{eqnarray}
for all $i=1,...,l$, with
$\ket{\psi_{\alpha}}=\sum_ic_\alpha^i\ket{i}$. This means that the number,
$D_\rho$, of the linear independent vectors of the set
$\{\ket{\uplambda_i}\}$ must satisfy
$D_\rho=l-\sum_\alpha (d_\alpha-1)\leq m-\sum_\alpha d_\alpha+\sum_\alpha 1=\sum_\alpha 1$. From Theorem 1 and the
definition of $\Delta\psi_{\alpha}\prec\Delta\varphi$, we can obtain that the number of
$c_{\alpha}^{i}\neq0$ is at least as many as that of $c_{i}\neq0$. Thus, there is
$D_\rho=\sum_\alpha 1\leq \frac mn$.

In passing, we would like to point that the phenomenon of bound coherence under strictly
incoherent operations was uncovered in Refs. \cite{Zhao,Lami,Lami1}
recently, i.e., there are coherent states from which no coherence can
be distilled via strictly incoherent operations in the asymptotic
regime. The necessary and sufficient condition for a state being bound
state was presented in Refs. \cite{Lami,Lami1}. Their result shows
that a state is a bound state if and only if it cannot contain any
rank-one submatrix. Comparing this result with the \emph{Corollary},
we obtain that, for any mixed state $\rho$, if we can transform it into a
pure coherent state $\ket {\varphi}$, then it cannot be a bound state.
However, in general, the converse is not true. Thus, the set of states
that can be transformed into a pure coherent state $\ket{\varphi}$ is a strictly
smaller set of the set of distillable states.

\emph{Conclusions.}--We have completed the operational task of
deterministic coherence distillation for a finite number of coherent
states under strictly incoherent operations. Specifically, we have
presented the necessary and sufficient condition for the
transformation from a mixed coherent state into a pure coherent state via
strictly incoherent operations, which recovers a connection between
the resource theory of coherence and the algebraic theory of majorization
lattice. With the help of this condition, we have presented the
deterministic coherence distillation scheme and we have
derived the maximum number of maximally coherent states that can be
obtained via this scheme.

\begin{acknowledgments}
We would like to thank Dian-Min Tong, Xiao-Dong Yu, and Qi-Ming Ding for thoroughly 
reading the manuscript, and for many suggestions, corrections, and comments,
 which have certainly helped to improve this paper. This work is
supported by NSF of China (Grant No.11775300), the
 National Key Research and Development Program of China
 (2016YFA0300603), and the Strategic Priority Research Program of
 Chinese Academy of Sciences No. XDB28000000.
\end{acknowledgments}

\end{document}